\documentclass[11pt,twoside]{article}

\usepackage{asp2014}
\usepackage{xcolor}

\aspSuppressVolSlug
\resetcounters

\begin{document}

\title{High performance visualization for Astronomy \& Cosmology: the VisIVO’s pathway toward Exascale systems}

\author{Eva~Sciacca,$^1$ Nicola~Tuccari,$^2$ Fabio~Vitello,$^1$ and Valentina~Cesare$^1$}
\affil{$^1$INAF Astrophysical Observatory of Catania, Catania, Italy; \email{eva.sciacca@inaf.it}}
\affil{$^2$Università di Catania, Dipartimento di Matematica e Informatica, Catania, Italy}

\paperauthor{Eva~Sciacca}{eva.sciacca@inaf.it}{0000-0002-5574-2787}{INAF}{Astrophysical Observatory of Catania}{Catania}{CT}{95123}{Italy}
\paperauthor{Nicola~Tuccari}{nicola.tuccari@inaf.it }{0009-0004-7802-2602}{Università di Catania}{Dipartimento di Matematica e Informatica}{Catania}{CT}{95123}{Italy}
\paperauthor{Fabio~Vitello}{fabio.vitello@inaf.it }{0000-0003-2203-3797}{INAF}{Astrophysical Observatory of Catania}{Catania}{CT}{95123}{Italy}
\paperauthor{Valentina~Cesare}{Author3Email@email.edu}{0000-0003-1119-4237}{INAF}{Astrophysical Observatory of Catania}{Catania}{CT}{95123}{Italy}



\begin{abstract}
Petabyte-scale data volumes are generated by observations and simulations in modern astronomy and astrophysics. Storage, access, and data analysis are significantly hampered by such data volumes and are leading to the development of a new generation of software tools. The Visualization Interface for the Virtual Observatory (VisIVO) has been designed, developed and maintained by INAF since 2005 to perform multi-dimensional data analysis and knowledge discovery in multivariate astrophysical datasets. Utilizing containerization and virtualization technologies, VisIVO has already been used to exploit distributed computing infrastructures including the European Open Science Cloud (EOSC).

We intend to adapt VisIVO solutions for high performance visualization of data generated on the (pre-)Exascale systems by HPC applications in Astrophysics and Cosmology (A\&C), including GADGET (GAlaxies with Dark matter and Gas) and PLUTO simulations, thanks to the collaboration within the SPACE Center of Excellence, the H2020 EUPEX Project, and the ICSC National Research Centre. In this work, we outline the evolution's course as well as the execution strategies designed to achieve the following goals: enhance the portability of the VisIVO modular applications and their resource requirements; foster reproducibility and maintainability; take advantage of a more flexible resource exploitation over heterogeneous HPC facilities; and, finally, minimize data-movement overheads and improve I/O performances.
\end{abstract}



\section{Introduction}

Modern astronomy and astrophysics produce massively large data volumes (in the order of petabytes) coming from observations or simulation codes executed on high performance supercomputers. Such data volumes pose significant challenges for storage, access and data analysis. Visual exploration of big datasets poses some critical challenges \cite{hey2009the} that must drive the development of a new generation of graphical software tools, specifically: (i) Interactivity to deal with datasets exceeding the local machine's memory capacity, so for complex visualizations the relevant computations should be performed close to the data to avoid time consuming streaming of large data volumes; (ii) Integration to be ideally fully integrated within the scientists' toolkit for seamless usage, abstracting from technical details related to the underlying high performance computing (HPC) resources, freeing scientists to concentrate in doing science; (iii) Collaboration to facilitate visualization, processing and analysis of big data in a collaborative manner within, e.g., science gateway technologies to allow collaborative activity between users and provide customization and scalability of data analysis/processing workflows, hiding underlying technicalities.

\section{Current Status}
INAF Astrophysical Observatory of Catania has been developing and maintaining the Visualization Interface for the Virtual Observatory (VisIVO) since 2005 \cite{gheller2006visivo} that has been recently extended with the ViaLactea Visual Analytic \cite{vitello2018vialactea} modules. VisIVO is developed adopting the Virtual Observatory standards and its main objective is to perform 3D and multi-dimensional data analysis and knowledge discovery of a-priori unknown relationships between multi-variate and complex astrophysical datasets.

VisIVO has already been deployed using Science Gateways \cite{becciani2015science} to access DCIs (including clusters, grids and clouds) using containerization and virtualization technologies, it has also been selected as one of the pilot applications deployed on EOSCpilot infrastructure demonstrating that the tools can be accessed using gateways and cloud platforms and it has been deployed on the European Open Science Cloud (EOSC), efficiently exploiting Cloud infrastructures and interactive notebooks applications \cite{sciacca2022scientific}.

\section{Peta-scale evolution direction}

During this period, and thanks to the collaboration within the SPACE Center of Excellence\footnote{SPACE CoE web page: \url{https://www.space-coe.eu/}} and the ICSC Italian National Research Centre for High Performance Computing, Big Data and Quantum Computing\footnote{ICSC web page: \url{https://www.supercomputing-icsc.it/}} we are planning to adapt VisIVO solutions for high performance visualisation, remote, in-situ, in-transit visualisation of data generated on the (pre-)exascale systems by HPC applications in Astrophysics and Cosmology (A\&C) including GADGET (GAlaxies with Dark matter and Gas intEracT) simulated data \cite{groth2023cosmological} and astrophysical fluid dynamics (PLUTO) simulations \cite{mignone2011pluto}.

In this work we present the preliminary related implementation activities tailored to pursue the following objectives: 1) Optimize and parallelize VisIVO modules to efficiently handle and process A\&C data on exascale computing resources; 2) Enhance the portability of the VisIVO modular applications and their resource requirements; 3) Foster reproducibility and maintainability; 4) Take advantage of a more flexible resource exploitation over heterogeneous HPC facilities (including also mixed HPC-Cloud resources); 5) Minimize data-movement overheads and improve I/O performances.

\section{Preliminary Results}

\subsection{Preliminary scalability tests}

VisIVO Importer parallel implementation was performed with a hybrid MPI-OpenMP approach and has been tested on the Pleiadi platform\footnote{\url{https://www.ict.inaf.it/computing/pleiadi/}}. Each Pleiadi node is equipped with two Intel(R) Xeon(R) CPU E5-2697 v4 @ 2.30GHz processors with BeeGFS parallel filesystem, allowing each node to access up to 36 cores. Tests were conducted on multiple configurations, ranging from one to four nodes and from one to 36 OpenMP threads per node. For each configuration, we considered one MPI process per node. Various datasets in the GADGET format with the following sizes were used: 4 GB, 16 GB, and 320 GB.

\begin{figure}[ht]
\centering
\includegraphics[width=0.8\textwidth]{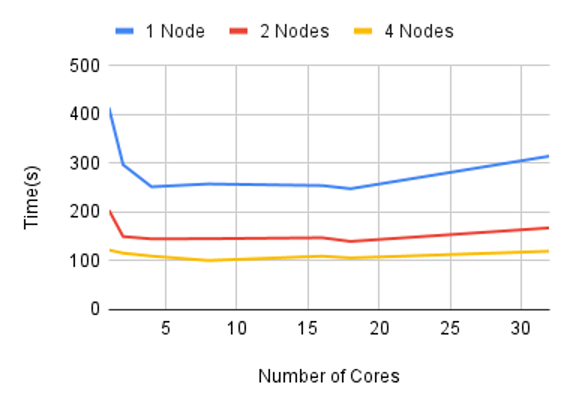}
\caption{VisIVO Importer execution time using 320 GB GADGET dataset.}
\label{fig:parallel320gb}
\end{figure}

The results in Figure~\ref{fig:parallel320gb} show that the execution time significantly decreases when the number of nodes increase. 

When working with the 320 GB dataset, the multithreading behavior remains consistent among one, two, and four nodes: it scales primarily up to 4 threads, and there are no significant speed gains observed when using more than 8 threads.

\subsection{Workflow abstractions}

We are investigating workflow abstractions to allow a portable representation of the VisIVO modular applications and their resource requirements, fostering reproducibility and maintainability, to take advantage of heterogeneous HPC facilities (including also mixed HPC-Cloud resources) while minimizing data-movement overheads. In particular, we would like to integrate VisIVO with StreamFlow\footnote{StreamFlow, \url{https://streamflow.di.unito.it/}} \cite{colonnelli2021streamflow} and Jupyter Workflow\footnote{\url{https://jupyter-workflow.di.unito.it/}} \cite{colonnelli2022jw}.

To test the workflow approach on VisIVO, we wrote a simple workflow with CWL, which executes a \texttt{VisIVOImporter} and \texttt{VisIVOViewer} command in sequence. This workflow was put on a GitHub repository and it was assigned the Apache 2.0\footnote{\url{https://spdx.org/licenses/Apache-2.0}} license.

\subsection{Interactive computing}
The Interactive Computing Service\footnote{ICS, \url{https://fenix-ri.eu/infrastructure/services/interactive-computing-services}} (ICS) will be exploited to explore the offered functionalities, in particular the ones related to the web interfaces enabling VisIVO pipelines. This work will include the implementation of Python wrappers to VisIVO Command Line Interfaces thus seamlessly integrating VisIVO with interactive notebooks and Python codes. Additionally, we will test and explore the Virtual Network Computing (VNC) based features for enhancing the capabilities of VLVA and, finally, we will test the StreamFlow integration within the service when it will be available.

\subsection{Fast I/O}
Additionally, we would like to investigate fast I/O techniques for optimizing the importing of large-scale datasets (currently employing MPI). For example, we would like to use the Cross-Application Programmable I/O (CAPIO)\footnote{CAPIO, \url{https://github.com/High-Performance-IO/capio}} \cite{martinelli2023capio}, to investigate
the integration of VisIVO workflows with the CAPIO middleware to boost its I/O performances without modifying the original codebase and allow it to coordinate the I/O within the VisIVO modules and, possibly, inject streaming capabilities into its workflow. 

\acknowledgements This work is funded by the European High Performance Computing Joint Undertaking (JU) and Belgium, Czech Republic, France, Germany, Greece, Italy, Norway, and Spain under grant agreement No 101093441 and it is supported by the spoke "FutureHPC \& BigData” of the ICSC – Centro Nazionale di Ricerca in High Performance Computing, Big Data and Quantum Computing – and hosting entity, funded by European Union – NextGenerationEU”.

\bibliography{C902}  


\end{document}